\input epsf
\magnification\magstephalf
\overfullrule 0pt
\def\gsim{\raise.3ex\hbox{$\;>$\kern-.75em\lower1ex\hbox{$\sim$}$\;$}}

\font\rfont=cmr10 at 10 true pt
\def\ref#1{$^{\hbox{\rfont {[#1]}}}$}


\font\fourteenbf=cmbx12 scaled\magstep1

\font\tenbfit=cmbxti10
\font\sevenbfit=cmbxti10 at 7pt
\font\fivebfit=cmbxti10 at 5pt
\newfam\bfitfam 
\textfont\bfitfam=\tenbfit  \scriptfont\bfitfam=\sevenbfit
\scriptscriptfont\bfitfam=\fivebfit

\font\fourteensy=cmsy10 scaled\magstep2
\font\fourteenmit=cmmi10 scaled\magstep2
\def\Bigmath{\textfont2=\fourteensy
             \textfont1=\fourteenmit
             \fourteenbf}

\font\tenbfit=cmbxti10
\font\sevenbfit=cmbxti10 at 7pt
\font\fivebfit=cmbxti10 at 5pt
\newfam\bfitfam 
\textfont\bfitfam=\tenbfit  \scriptfont\bfitfam=\sevenbfit
\scriptscriptfont\bfitfam=\fivebfit

\font\tenbit=cmmib10
\newfam\bitfam
\textfont\bitfam=\tenbit%

\font\tenmbf=cmbx10
\font\sevenmbf=cmbx7
\font\fivembf=cmbx5
\newfam\mbffam
\textfont\mbffam=\tenmbf \scriptfont\mbffam=\sevenmbf
\scriptscriptfont\mbffam=\fivembf

\font\tenbsy=cmbsy10
\newfam\bsyfam 
\textfont\bsyfam=\tenbsy%

 \def\b {\beta}  
\def\e{\epsilon}

\def\pmb#1{\setbox0=\hbox{#1}
 \kern.05em\copy0\kern-\wd0 \kern-.025em\raise.0433em\box0 }

\def\slash{/\kern-.5em}

\def \half {{\textstyle {1 \over 2}}}

\def\b{$\bullet~$}

\def\boxit#1{\vbox{\hrule\hbox{\vrule\kern1pt\vbox
{\kern1pt#1\kern1pt}\kern1pt\vrule}\hrule}}

\def\h{\hfill\break}
\parskip=6pt
\parindent=0pt
\hsize=17truecm\hoffset=-5truemm
\vsize=23truecm
\def\footnoterule{\kern-3pt
\hrule width 17truecm \kern 2.6pt}


\catcode`\@=11 

\def\nolabels{\def\wrlabeL##1{}\def\eqlabeL##1{}\def\reflabeL##1{}}
\def\writelabels{\def\wrlabeL##1{\leavevmode\vadjust{\rlap{\smash%
{\line{{\escapechar=` \hfill\rlap{\sevenrm\hskip.03in\string##1}}}}}}}%
\def\eqlabeL##1{{\escapechar-1\rlap{\sevenrm\hskip.05in\string##1}}}%
\def\reflabeL##1{\noexpand\llap{\noexpand\sevenrm\string\string\string##1}}}
\nolabels
\global\newcount\refno \global\refno=1
\newwrite\rfile
\def\defref{$^{{\hbox{\rfont [\the\refno]}}}$\nref}
\def\nref#1{\xdef#1{\the\refno}\writedef{#1\leftbracket#1}%
\ifnum\refno=1\immediate\openout\rfile=refs.tmp\fi
\global\advance\refno by1\chardef\wfile=\rfile\immediate
\write\rfile{\noexpand\item{#1\ }\reflabeL{#1\hskip.31in}\pctsign}\findarg}
\def\findarg#1#{\begingroup\obeylines\newlinechar=`\^^M\pass@rg}
{\obeylines\gdef\pass@rg#1{\writ@line\relax #1^^M\hbox{}^^M}%
\gdef\writ@line#1^^M{\expandafter\toks0\expandafter{\striprel@x #1}%
\edef\next{\the\toks0}\ifx\next\em@rk\let\next=\endgroup\else\ifx\next\empty%
\else\immediate\write\wfile{\the\toks0}\fi\let\next=\writ@line\fi\next\relax}}
\def\striprel@x#1{} \def\em@rk{\hbox{}} 
\def\lref{\begingroup\obeylines\lr@f}
\def\lr@f#1#2{\gdef#1{\defref#1{#2}}\endgroup\unskip}
\newwrite\lfile
{\escapechar-1\xdef\pctsign{\string\%}\xdef\leftbracket{\string\{}
\xdef\rightbracket{\string\}}}

\def\writestop{\def\writestoppt{\immediate\write\lfile{\string\p
ageno%
\the\pageno\string\startrefs\leftbracket\the\refno\rightbracket%
\string\def\string\secsym\leftbracket\secsym\rightbracket%
\string\secno\the\secno\string\meqno\the\meqno}\immediate\closeout\lfile}}
\def\writestoppt{}\def\writedef#1{}
\catcode`\@=12 
\rightline{MAN/HEP/2013/19}
\rightline{DAMTP-2013-38}
\bigskip
\centerline{\fourteenbf \Bigmath{$pp$} and \Bigmath{$\bar pp$} total cross sections and elastic scattering}
\vskip 8pt
\centerline{A Donnachie}
\centerline{School of Physics and Astromony, University of Manchester}
\vskip 5pt
\centerline{P V Landshoff}
\centerline{DAMTP, Cambridge University$^*$}
\footnote{}{$^*$ email addresses: Sandy.Donnachie@hep.manchester.ac.uk, \ 
pvl@damtp.cam.ac.uk}
\bigskip
{\bf Abstract}

It is shown that $p p$ and $p\bar p$ data, including those from the TOTEM experiment, agree well with Regge theory.
\bigskip
{\bf 1 Introduction}

In this paper we analyse whether the highly-accurate elastic scattering and 
total cross section data now available\defref\totem
{
TOTEM collaboration: 
G Antchev et al, Europhysics Letters 101 (2013) 21002 and 21004,
Physical Review Letters 111 (2013) 012001
}
from the LHC help to shed any new light on the unanswered questions
about the theory.

The only theory that we have is Regge theory\defref\regge
{
T Regge, Il Nuovo Cimento, 14 (1959) 951; G F Chew and S C Frautschi,
Physical Review Letters 8 (1962) 41
}. While it has been hugely successful\defref\collins
{
P D B Collins, {\it Introduction to Regge theory}, Cambridge University Press 
(1977)
}, 
our understanding of it has a significant 
gap: we know how to describe the exchanges of single particles,  but
even after five decades we do not know how to calculate double and higher
exchanges. Further, to describe the data we need to introduce an exchange
that is not obviously associated with particle exchange. Called pomeron
exchange, it is possible that it corresponds to glueball exchange.

According to Regge theory, particle exchange contributes simple power
behaviour $ s^{\epsilon}$ to total cross sections. In our original
analysis three decades ago\defref\sigtot{
A Donnachie and P V Landshoff, Physics Letters B296 (1992) 227
}
of the data then available, up to $\sqrt s=62$~GeV,
we introduced just two powers to fit all hadron-hadron total cross sections,
$\epsilon_1$ close to 0.08 corresponding to pomeron exchange and 
$\epsilon_R$ close to $\half$ corresponding to the nearly exchange-degenerate
$\rho,\omega,f_2,a_2$ exchange. In the light of the data subsequently
obtained up to Tevatron energies, Cudell and collaborators\defref\cudell
{
J R Cudell, K Kang and S K Kim Physics Letters B395 (1997) 311
}
concluded that the value of $\epsilon_1$ was somewhat larger, close to
0.096. 

When deep inelastic $ep$ scattering data at small $x$ became 
available\defref\hera
{
H1 and ZEUS collaborations: F D Aaron et al,  JHEP 1001 (2010) 109 
} 
from HERA, we showed that these were most simply described by introducing
a second pomeron, the hard pomeron\defref\hardpom
{
A Donnachie and P V Landshoff, Physics Letters B437 (1998) 408
},
and that by doing so we could put the conventional DGLAP evolution analysis
 at small $x$ on a sounder footing\defref\dglap
{
A Donnachie and P V Landshoff, Physics Letters B533 (2002) 277
}.
We then pointed out\defref\factor
{
A Donnachie and P V Landshoff, Physics Letters B595 (2004) 393
}
that a corresponding power behaviour $s^{\epsilon_0}$ with
$\epsilon_0$ close to 0.4 might well be present in hadron-hadron total
cross sections.

Through the optical theorem, total cross sections are closely linked to elastic
scattering, and so fitting one and ignoring the other does not make sense.
A particular challenge to fitting  elastic scattering data is the
remarkable dip structure first discovered\defref\chhav
{
CHHAV collaboration: E Nagy et al, Nuclear Physics B150 (1979) 221
}
in $pp$ elastic scattering at the CERN ISR. We predicted\defref\elastic
{
A Donnachie and P V Landshoff, Phys Lett 123B (1983) 345
}
that this dip would be filled in for $\bar pp$ scattering, and this
was subsequently confirmed\defref\pbarp
{
CERN-R-420 collaboration: A Breakstone et al, Physical Review Letters 54 (1985)
2180
}
though, as we again find in this paper, a detailed correct description of 
the $\bar pp$ data is very difficult to achieve.

Reproducing a dip requires the
simultaneous near-vanishing of both the real and imaginary parts of
the amplitude, so at least three terms need to be involved. The energies at which 
the dip is seen are quite large, so the contributions from
$\rho,\omega,f_2,a_2$ exchange are too small, and to cancel the imaginary
part of single pomeron exchange $P$ we need also two-pomeron exchange $PP$. 
To cancel the real part we bring in triple-gluon exchange, since this
appears\defref\ggg{
A Donnachie and P V Landshoff, Zeits Physik C2 (1979) 55
}
to dominate the elastic amplitude at large values of $t$, giving it
an energy-independent behaviour $t^{-4}$.

Two-pomeron exchange is, of course, important for another reason. Without
it, the forward amplitude would grow so large with increasing energy that
unitarity would be violated. For this reason, in our original fit\ref{\sigtot}
we emphasised that the power $\epsilon_1$ was only an effective
power, valid over a limited range of energies. The term $PP$ is negative
and so helps to avoid such a breakdown of unitarity, but its magnitude
increases more rapidly than that of the term $P$, so that to prevent the
total cross section from becoming negative
it will ultimately be necessary to include also a further term
$PPP$, and so on. But there is no indication that this is needed at
presently-accessible energies. We shall just work with $P+PP+ggg$,
together with $\rho,\omega,f_2,a_2$ exchange.

In this, our approach is different from that of most other
authors, who bring in the complete series $P+PP+PPP+PPPP+\dots$. 
To calculate this, they use an eikonal formula, but this has little
or no theoretical justification.
As we have stated above, we do not have the theoretical knowledge
correctly to calculate any of the terms beyond single exchange $P$,
though we do have some knowledge of their general properties, which 
enables one to make models.

A conclusion of this paper is that, while the data for $pp$ and
$\bar pp$ elastic scattering and total cross sections do not exclude
a small contribution from the exchange of the hard pomeron, it is not
needed to fit the data. And an interesting feature of what we find 
below is that, even though
the term $PP$ behaves as combination of a power of $s$ and log $s$, 
our fit to the data makes the combination $P+PP$ effectively behave 
as a simple power of $s^{\epsilon_1}$ over a very wide range of energy, 
up the highest values that are likely ever to be accessible. We find that 
this effective power $\epsilon_1$ is close to the value 0.096 arrived at 
in reference [{\cudell}]. Contrary to what we concluded in a previous
paper\defref\previous
{
A Donnachie and P V Landshoff, arxiv.org:1112.2485 
},
the hard pomeron is not needed to give a good fit to the data, though it may be present at some level.

\bigskip

{\bf 2 Fit to the data}

In our original fit\ref{\sigtot}  to the total cross section data
we took all four ``reggeon'' trajectories $\rho,\omega,f_2,a_2$ to be 
degenerate, so that their exchanges all contributed the same power of
$s$. However, a Chew-Frautschi plot of particle masses
shows that this degeneracy is far from exact -- see figure 2.13
of our book\defref\book
{
A Donnachie, H G Dosch, P V Landshoff and O Nachtmann, {\sl Pomeron Physics
and QCD}, Cambridge University Press (2002)
},
which suggests that it is more accurate to work with
two pairs of degenerate trajectories, $f_2,a_2$ with slope 
0.8 GeV$^{-2}$ and intercept $1+\epsilon_+$ close to 0.7, and $\rho,\omega$
with slope 0.92 GeV$^{-2}$ and intercept $1+\epsilon_-$ close to 0.5.  

As we have explained, we find that the hard pomeron
is not needed to fit the data, so we introduce three trajectories
$$
\alpha_i(t)=1+\epsilon_i+\alpha_i't~~~~i=  P,\pm
\eqno(1a)
$$
and we have three single-exchange terms:
$$
A(s,t)=
-{X_PF_P(t)\over 2\nu}~e^{-\half i\pi \alpha_P(t)}~(2\nu \alpha_P')^ {\alpha_{P}(t)} 
-{X_+F_+(t)\over 2\nu}~e^{-\half i\pi \alpha_+(t)}~(2\nu \alpha_+')^ {\alpha_+(t)} $$$$
~~~~~~~~~~~~~~~~~~\mp {iX_-F_-(t)\over 2\nu}~e^{-\half i\pi \alpha_-(t)}~(2\nu \alpha_-')^ {\alpha_-(t)} 
\eqno(1b)
$$
for the $pp/\bar pp$ amplitude. The normalisation is such that
$\sigma^{\hbox{{\sevenrm TOT}}}(s)=$ Im $A(s,0)$. The form factors $F_i(t)$ 
are unknown. We find that a good fit is obtained by taking them all to be the 
same and of the simple form
$$
F(t)=A e^{at}+(1-A)e^{bt}
\eqno(1c)
$$

In our fit, we will assume that we can neglect the simultaneous exchange
of two reggeons, $RR$, and the exchange of a reggeon together with the
pomeron, $RP$. So $\epsilon_{\pm}$ should be regarded
as effective powers. We shall find that the best fit to the data gives
values for them close to those from the Chew-Frautschi plots quoted above, 
so justifying our assumption that the exchanges $RR$ and $RP$ are small.

As we have said, we do not know how to calculate the term $PP$. What is 
known\ref{\book} is that it corresponds to a trajectory
$$
\alpha_{PP}(t)=1+2\epsilon_P+\half\alpha'_Pt
\eqno(2a)
$$
but that the contribution to the amplitude behaves not just as the simple 
power $s^{\alpha_{PP}(t)}$; there are additional logarithmic factors in the 
denominator. Also, the normalisation is unknown. Our procedure was to start 
with the eikonal form and adapt it until we achieved a good fit. Numerous 
variations of the eikonal double-exchange form were tried, none of which were
ideal. This led us to a $PP$ contribution of the form 
$$
{X_P^2\over 64\pi\nu}~e^{-\half i\pi \alpha_{PP}(t)}~(2\nu \alpha_{P}')^{\alpha_{PP}(t)}
\Big[{A^2\over a/\alpha_{P}'+L}~ 
e^{\half at}+
{(1-A)^2\over b/\alpha_{P}'+L}~ 
e^{\half bt}\Big]
$$$$
L=\log(2\nu\alpha_{P}')-\half i\pi
\eqno(2b)
$$
This contains the key ingredients of the eikonal, namely the trajectory, the
logarithmic factor in the denominator and the modification of the argument of 
the exponentials in the form factor. It differs from the eikonal form only 
in that it does not include a cross term involving $A(1-A)$. Omitting this 
term very significantly improves the fit. This should not be a surprise since,
as we have said, the eikonal has no theoretical justification. Indeed, it is 
quite surprising that this simple modification to it works so well. 
{\topinsert
\centerline{\epsfxsize=0.48\hsize\epsfbox[85 570 320 760]{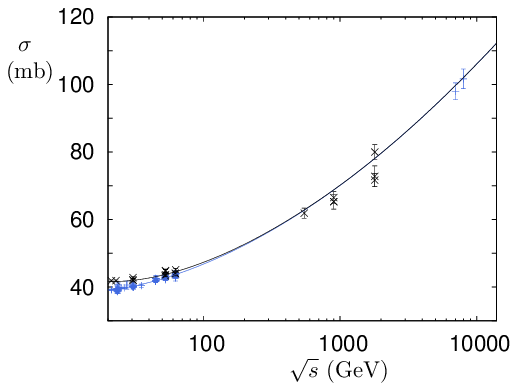}}

Figure 1: Fit to data for the total cross sections for $pp$ and $\bar pp$
scattering. The data are taken from the PDG compilation\defref\pdg
{
Particle Data Group: J Beringer et al, Physical Review D86 (2012) 010001,  pdg.lbl.gov/2012/hadronic-xsections/
}. 
The Tevatron $\bar pp$ points are not included in the fit.
\endinsert
\pageinsert
\epsfxsize=0.48\hsize\epsfbox[80 60 390 290]{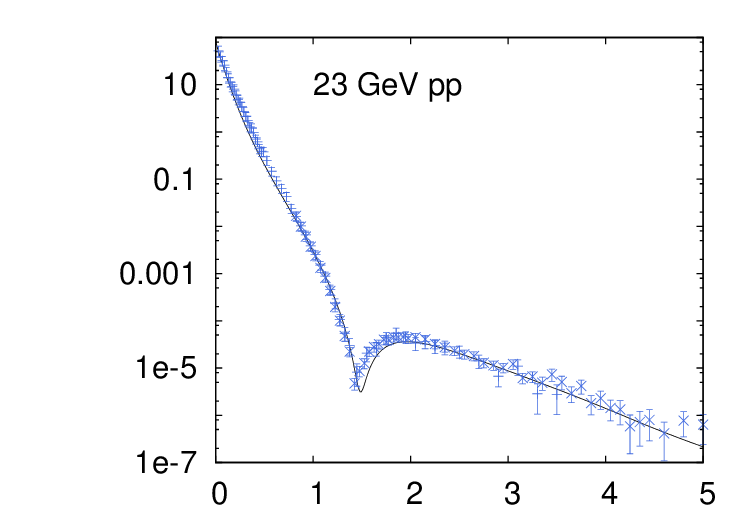}\hfill
\epsfxsize=0.48\hsize\epsfbox[80 60 390 290]{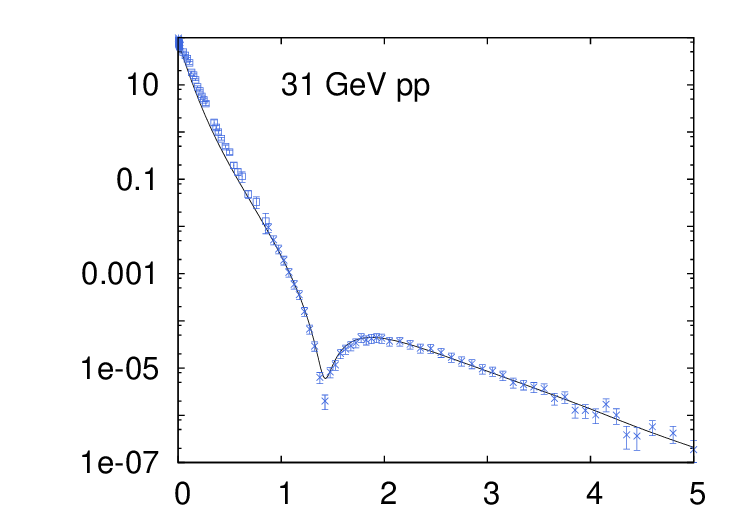}

\epsfxsize=0.48\hsize\epsfbox[80 60 390 290]{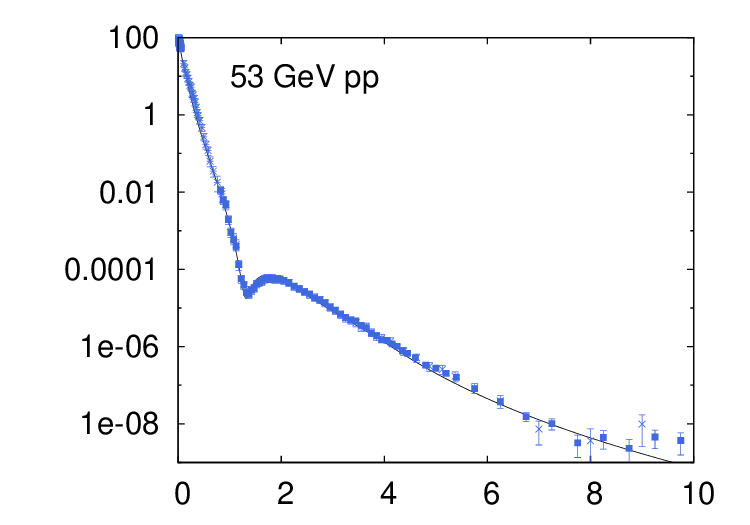}\hfill
\epsfxsize=0.48\hsize\epsfbox[80 60 390 290]{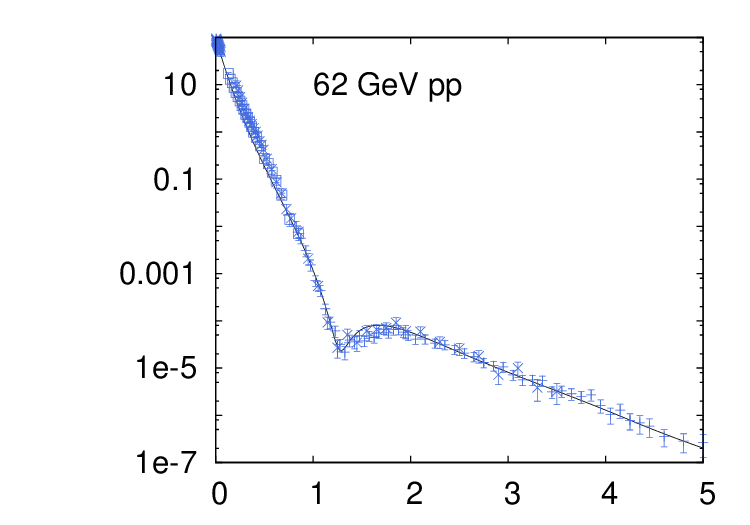}

\epsfxsize=0.48\hsize\epsfbox[80 60 390 290]{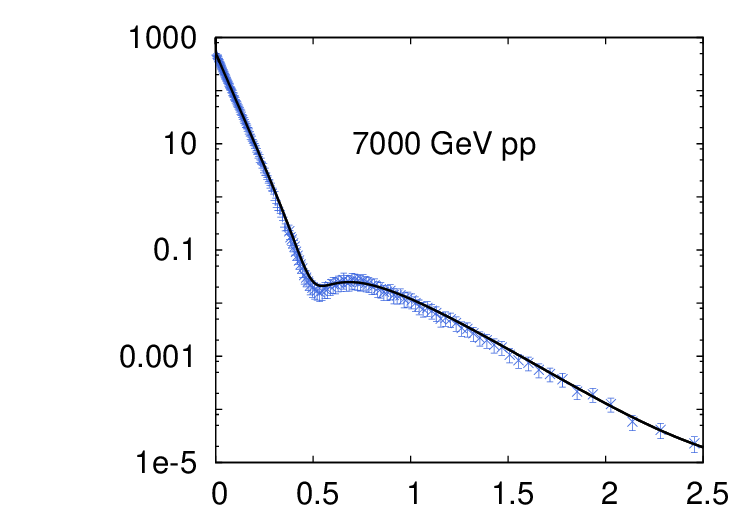}\hfill
\epsfxsize=0.48\hsize\epsfbox[80 60 390 290]{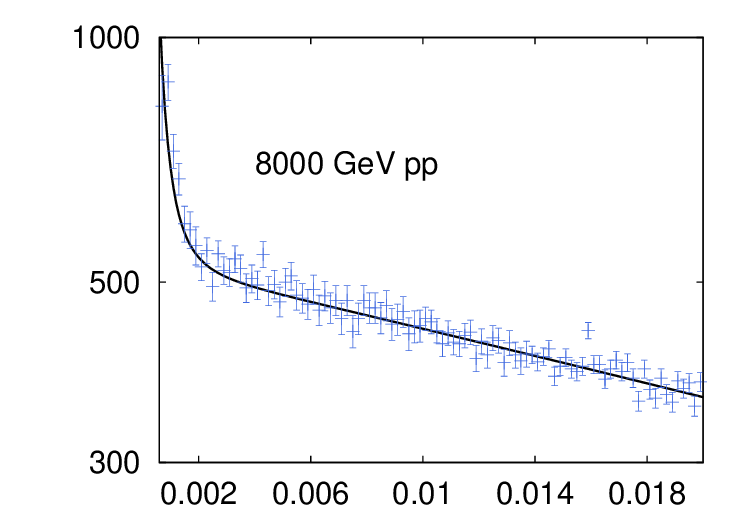}

Figure 2: Fits to data\ref\totem\ref\chhav\defref\Kwak{Kwak et al, Physics Letters 58B 
(1975) 233}\defref\Breakstone{CERN-R-420 collaboration: A Breakstone et al, 
Nuclear Physics B248 (1984) 253}
for $pp$ elastic scattering: $d\sigma/dt$ in 
mb~GeV$^{-2}$ plotted against $t$ in GeV$^2$.

\endinsert}

{\topinsert

\epsfxsize=0.48\hsize\epsfbox[80 60 390 290]{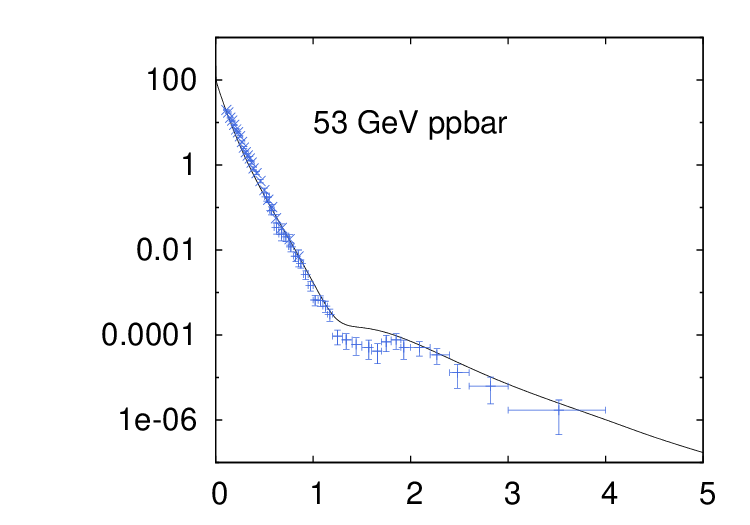}\hfill
\epsfxsize=0.48\hsize\epsfbox[80 60 390 290]{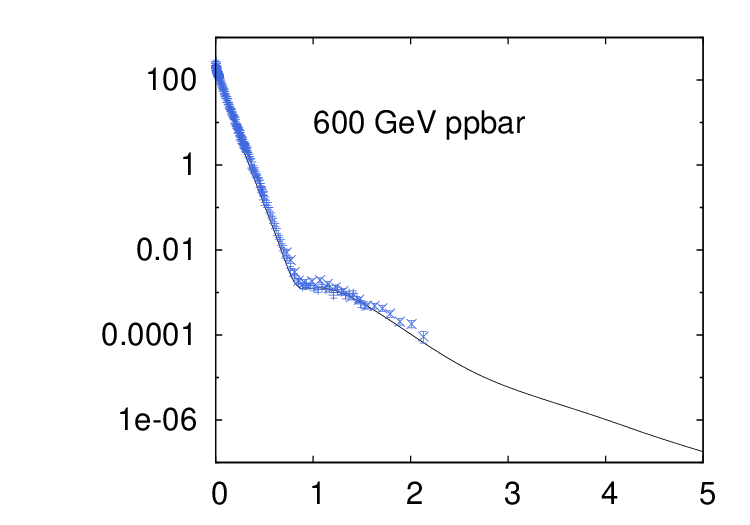}\hfill
\medskip
\centerline{\epsfxsize=0.48\hsize\epsfbox[80 60 390 290]{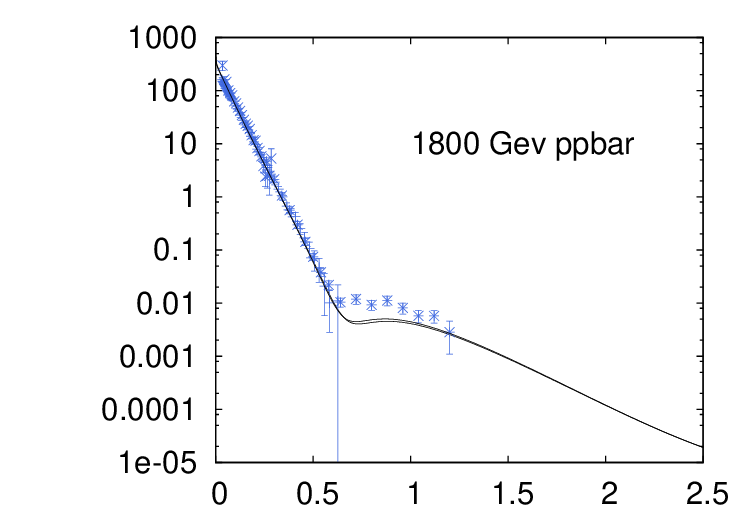}}
Figure 3: fits to data\ref\pbarp\ref\Breakstone
\defref\Battiston{UA4 collaboration: R Battiston et al, Physics Letters 127B 
(1983) 472\h
{UA1 collaboration: G Arnison et al, Physics Letters 128B (1983) 
336}\h
{UA4 collaboration: M Bozzo et al, Physics Letters 155B (1985) 197}
\h
{UA4 collaboration: D Bernard et al, Physics Letters 171B (1986) 142 and 
198B (1987) 583}\h
{E710 collaboration: N Amos et al, Physics Letters 247B (1990) 
127}\h
{CDF collaboration: F Abe et al, Physical Review D50 (1994) 5518}
\h
E710 collaboration: V M Abazov et al, Physical Review D86 (2012) 012009}
for $\bar pp$ elastic scattering: $d\sigma/dt$ in 
mb~GeV$^{-2}$ plotted against $t$ in GeV$^2$.

\endinsert}

We must also include the term $ggg$ corresponding to triple-gluon exchange.
At large $t$, say for $|t|>t_0$, it behaves as\defref\ggg
{
A Donnachie and P V Landshoff, Zeits. Physik C2 (1979) 55
}
$$
g(t)=C~{t_0^3\over t^4} ~~~~~|t|>t_0
\eqno(3a)
$$
For $|t|<t_0$ we fit this smoothly on to some function that does not diverge
as $t\to 0$. By trial and error we arrived at
$$
g(t)={C\over t_0}e^{2(1-t^2/t_0^2)}
\eqno(3b)
$$

Finally, in order to fit elastic scattering data at very small $t$, we include 
in the amplitudes the Coulomb term
$$
\mp {8\pi\alpha_{\hbox{\sevenrm EM}}\over t}
\eqno(4)
$$

We perform a least $\chi^2$ fit simultaneously to $pp$ and $\bar pp$ elastic 
scattering data down to centre-of-mass energy 23~ GeV, and total cross section
data down to 6~GeV. We include the very-small-$t$ preliminary TOTEM data, but 
not those at lower energies. This gives the parameter set:
$$
\matrix{
\e_P=0.110~~~\e_+=-0.327~~~\e_-=-0.505~~~~X_P= 339~~~ X_+= 212~~~ X_-= 104~~~~
\alpha'_P=0.165~\hbox{GeV}^{-2} \cr
A=0.682~~~ a=7.854~\hbox{GeV}^{-2}~~~b= 2.470~\hbox{GeV}^{-2}\cr
C=0.0406~~~t_0= 4.230~\hbox{GeV}^2 \cr} 
\eqno(5)
$$
\bigskip
{\bf 3 Results}

Figure 1 shows our fit to data for the $pp$ and $\bar pp$ total cross sections.
 We did not use the conflicting data from the Tevatron: the E710
measurement\defref\e710
{
E710 collaboration: N Amos et al, Physical Review Letters 68 (1992) 2433
}
is significantly below that of CDF\defref\cdf
{
CDF collaboration: F Abe et al, Physical Review D50 (1994) 5550
}
and it will be seen that our fit favours the latter.

Figure 2 shows the $pp$ elastic scattering data that we used, compared with 
the fit. If we had not included the preliminary 8~TeV data\defref\turini
{
TOTEM collaboration, talk by N Turini, http://totem.web.cern.ch/Totem/conferences/conf{$\underline{~}$}tab2013.html
} at very small $t$ the
fit would have passed just below them, while having little effect on the 
other plots. 
Figure 3 shows the data we used for elastic $\bar pp$ scattering, 
compared with the fit. 

Our fit did not use the very-small-$t$ data shown in figure 4. However when 
we include the Coulomb term (4) we predict these data surprisingly well.

{\pageinsert
\epsfxsize=0.48\hsize\epsfbox[80 60 390 290]{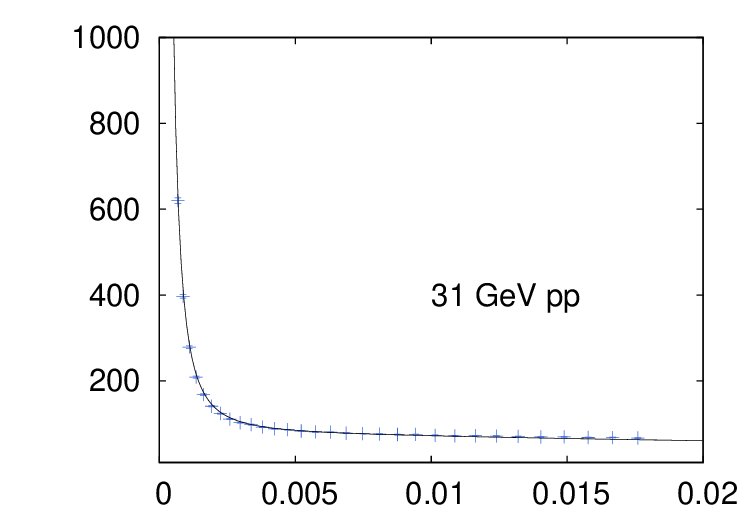}\hfill
\epsfxsize=0.48\hsize\epsfbox[80 60 390 290]{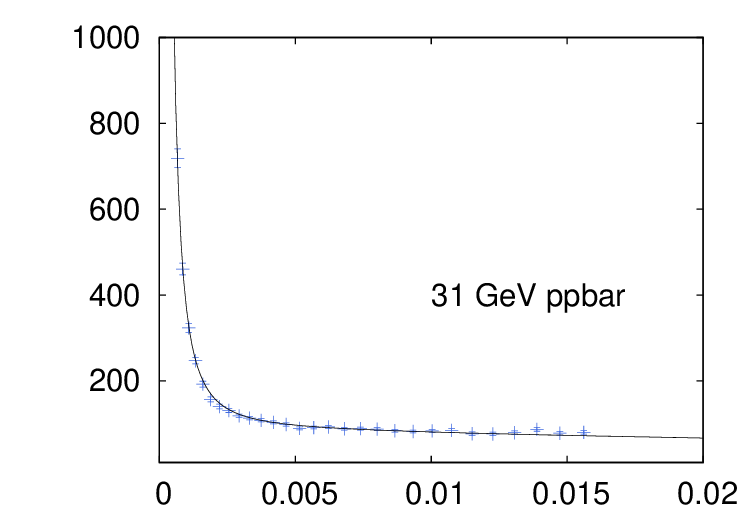}

\epsfxsize=0.48\hsize\epsfbox[80 60 390 290]{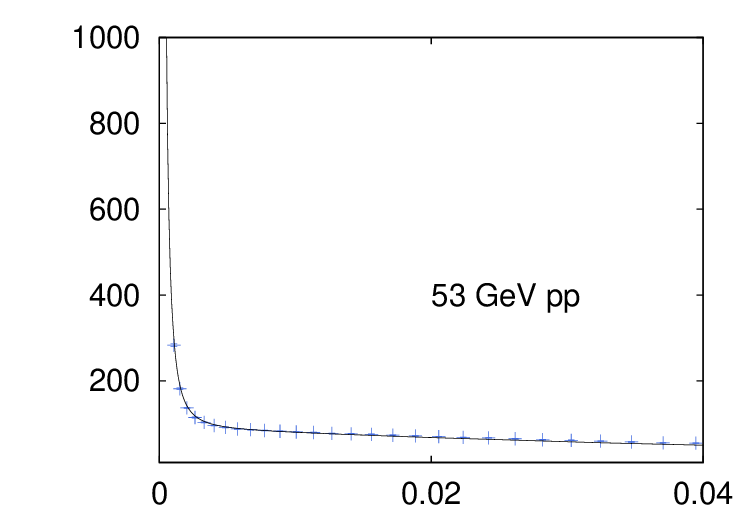}\hfill
\epsfxsize=0.48\hsize\epsfbox[80 60 390 290]{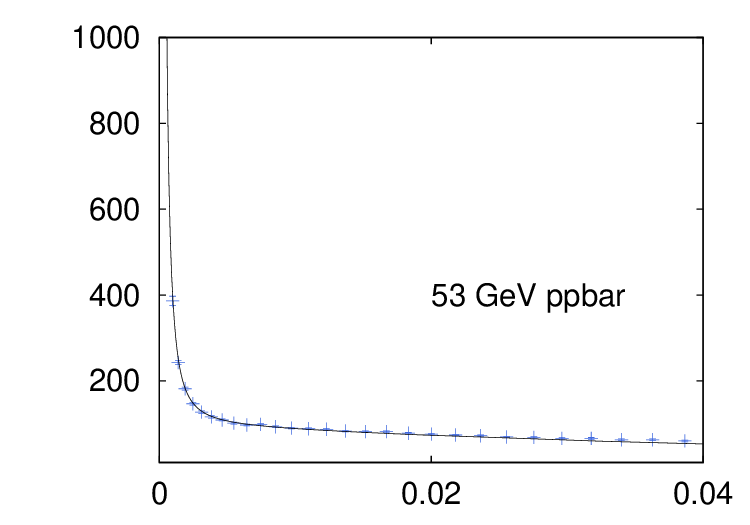}

Figure 4: Comparison with data at very small $t$: $d\sigma/dt$ in mb~GeV$^{-2}$ plotted 
against $t$ in GeV$^2$. The data are from \defref\AmosC{R-211 collaboration:
N Amos et al, Nuclear Physics B262 (1985) 689}
\bigskip
\epsfxsize=0.48\hsize\epsfbox[85 570 320 760]{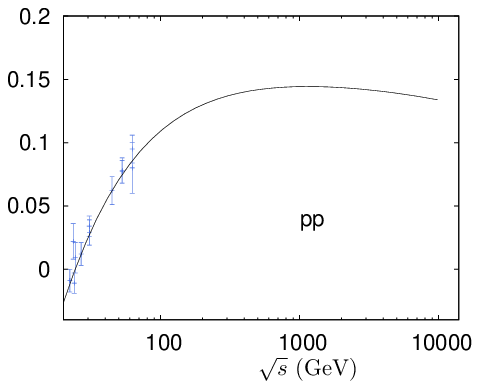}\hfill
\epsfxsize=0.48\hsize\epsfbox[85 570 320 760]{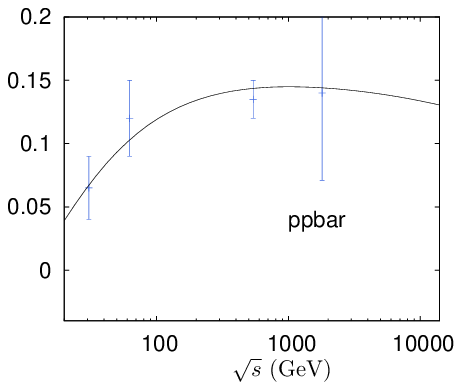}

Figure 5: Comparison with data\ref\AmosC 
\defref\Gross{E381 collaboration: D Gross, Physical Review Letters 41 (1978) 217\h
{FNAL-069A collaboration: L A Fajardo, Physical Review D24 (1981) 46}\h
{UA6 collaboration: R E Breedon, Physics Letters 216B (1989) 459}\h 
{E710 collaboration: N Amos et al, Physical Review Letters 68 
(1992) 2433}\h
{UA4 collaboration: C Augier et al, Physics Letters 316B (1993) 448}\h
R-211 collaboration, N Amos et al, Physics Letters 128B (1983) 343\h
U Amaldi et al, Physics Letters B66 (1977) 390}
for the ratio of the real to imaginary parts of the forward 
$pp$ and $\bar pp$ amplitudes. 
\endinsert}
\bigskip
\topinsert
\centerline{\epsfxsize=0.48\hsize\epsfbox[70 580 310 760]{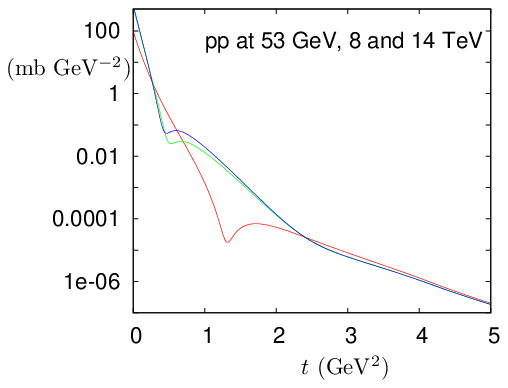}}

Figure 6: The $pp$ differential cross section at three energies
\bigskip
\bigskip
\centerline{\epsfxsize=0.48\hsize\epsfbox[85 570 320 760]{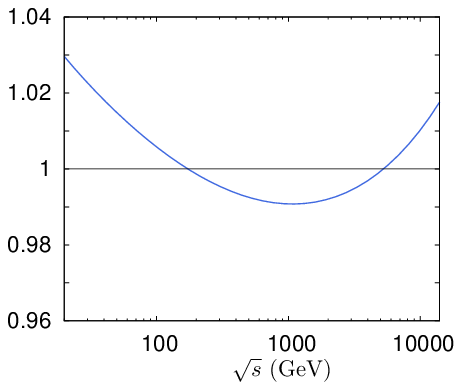}}

Figure 7: Ratio of the simple power $18.23~s^{0.0958}$ to $(P+PP)$
\endinsert

{\bf 4 Conclusions}

Our model makes various simplifications, but describes the data for
$pp$ scattering well and for $\bar pp$ scattering quite well, over a wide range 
of energy. Our simplifications include:
\midinsert\leftskip 1 truecm
\b making the three form factors $F_i(t)$ identical and of the simple form (1c)\h
\b making the trajectories $\alpha_i(t)$ linear\h
\b making the $\rho$ and $\omega$ trajectories degenerate, and also
the $f_2$ and $a_2$\h
\b omitting all non-single exchanges other than $PP$ and taking it to have the
simple form (2b)
\b assuming the simple form (3b) for the $ggg$ term when $t$ is not large.
\endinsert

Figure 4 shows an example of data that were not used to make the fit but
are described well by it. Another such example is the ratio of the real to
imaginary parts of the forward amplitudes, figure 5.

A correct description of the dips 
is challenging and our simple model is able to describe those in $pp$ 
scattering rather better than in $\bar pp$ scattering. 
We will
not succumb to the temptation to say that, having been taken somewhat
hurriedly in the very last few days of operation of the CERN Instersecting
Storage Rings, the $\bar pp$ data at 53~GeV are unreliable.

The triple-gluon-exchange term $g(t)$ plays a 
key role in giving the dips. At large enough $t$ it results in
$d\sigma/dt\sim 0.073/t^8$, somewhat smaller than our old fit\ref{\ggg}. 
Figure 6 shows the $pp$ elastic differential cross section at various energies.
The data make our fit very energy-independent for 
$|t|>4$ GeV$^2$, where it is dominated by the term $ggg$. We
have previously\defref\interest
{
A Donnachie and P V Landshoff,  Physics Letters B387 (1996) 637
}
drawn attention to the interest of checking whether, at sufficiently high 
energy, this energy independence might give way to a steady increase with energy.

The term $P$ contributes a behaviour $s^{0.110}$ to the total cross sections, while $PP$ is negative and behaves as 
$s^{0.220}$ together with the denominator logarithmic factors shown in (2b). 
Figure 7 shows that, over a very wide range of values of $\sqrt s$,
together their behaviour is very close to the simple power behaviour
$s^{0.096}$ that was extracted from the data by Cudell and 
collaborators\ref{\cudell}. The Froissart-Lukaszuk-Martin bound\defref\flm
{
M~Froissart,  Physical Review {123} (1961) 1053;
L~Lukaszuk and A~Martin,  Il Nuovo Cimento { 47A} (1967) 265
}
is about 20~barns at LHC energies and so has no relevance: our fit
confirms that asymptopia is a really long way away\defref\menon
{
M J  Menon and P V R G  Silva, arXiv:1305.2947
}.

Although a satisfactory fit can also be obtained with the hard pomeron 
included, our results show that it is not necessary.

\vfill\eject
\medskip\immediate\closeout\rfile\writestoppt
\baselineskip=13pt{{\bf References}}\medskip{\frenchspacing%
\parindent=20pt\escapechar=` \input refs.tmp\medskip}\nonfrenchspacing
\bye